\documentclass[prx,reprint,twocolumn,amsmath,superscriptaddress,longbibliography]{revtex4-1}
\usepackage{amsmath,amssymb,bm,mathrsfs,graphicx, braket, times,amsthm,enumerate, scrextend}
\usepackage[colorlinks=true,citecolor=blue,linkcolor=blue]{hyperref}
\usepackage{longtable}
\usepackage{multirow}
\usepackage{array}
\usepackage{bigstrut}
\usepackage[all,cmtip]{xy}
\usepackage[normalem]{ulem}
\usepackage[usenames,dvipsnames]{color}
\usepackage{makecell}
\usepackage{xcolor}

\begin{document}
\title{All hourglass bosonic excitations in the 1651 magnetic space groups and 528 magnetic layer groups}
\author{Dongze Fan}
\affiliation{National Laboratory of Solid State Microstructures and School of Physics, Nanjing University, Nanjing 210093, China}
\affiliation{Collaborative Innovation Center of Advanced Microstructures, Nanjing University, Nanjing 210093, China}
\author{Xiangang Wan}
\affiliation{National Laboratory of Solid State Microstructures and School of Physics, Nanjing University, Nanjing 210093, China}
\affiliation{Collaborative Innovation Center of Advanced Microstructures, Nanjing University, Nanjing 210093, China}
\affiliation{International Quantum Academy, Shenzhen 518048, China.}
\author{Feng Tang}\email{fengtang@nju.edu.cn}
\affiliation{National Laboratory of Solid State Microstructures and School of Physics, Nanjing University, Nanjing 210093, China}
\affiliation{Collaborative Innovation Center of Advanced Microstructures, Nanjing University, Nanjing 210093, China}
\begin{abstract}
The band connectivity as imposed by the compatibility relations between the irreducible representations of little groups can give rise to the exotic hourglass-like shape composed of four branches of bands and five band crossings (BCs). Such an hourglass band connectivity could enforce the emergence of nontrivial excitations like Weyl fermion, Dirac fermion or even beyond them.  On the other hand, the bosons, like phonons, magnons, and photons, were also shown to possess nontrivial topology and a comprehensive symmetry classification of the hourglass bosonic excitations would be of great significance to both materials design and device applications.  Here we firstly list all concrete positions and representations of little groups in the Brillouin zone (BZ) related with the hourglass bosonic excitations in all the 1651 magnetic space groups and 528 magnetic layer groups, applicable to three dimensional (3D) and two dimensional (2D) systems, respectively. 255 (42) MSGs (MLGs) are found to essentially host such hourglass BCs: Here ``essentially'' means that the bosonic hourglass BC exists definitely as long as the studied system is crystallized in the corresponding MSG/MLG. We also perform first-principles calculations on hundreds of 3D nonmagnetic materials essentially hosting hourglass phonons and propose that the 2D material AlI can host hourglass phonons. We choose AuX (X=Br and I) as illustrative examples to demonstrate that two essential hourglass band structures  can coexist in the phonon spectra for both materials while for AuBr,  an accidental band crossing sticking two hourglasses is found interestingly. Our results of symmetry conditions for hourglass bosonic excitations can provide a useful guide of designing artificial structures with hourglass bosonic excitations. %, especially for future studies on those with time-reversal symmetry breaking.
\end{abstract}
\maketitle
\date{\today}

%\section{Introduction}\label{intro}
\section{Introduction}

 Various topological band crossings (BCs) in bulk and boundary electronic systems have attracted extensive interest in the past nearly two decades \cite{Weyl-Wan, Balatsky, RMP-AV,RMP-LV,RMP-kane,RMP-qi,RMP-bansil,RMP-chiu,nature-review-mag} on which crystallographic symmetries \cite{Bradley, Bilbao, Dresselhaus} could impose diverse constraints. On the other hand, bosons, such as phonons, photons, and magnons, emergent in crystalline materials or artificial structures, have also been proposed to carry nontrivial band topology \cite{photonics-review,phonon-review,magnons-review}. Topological bosons are expected to be endowed with novel consequences like novel quantum information storage and processing \cite{photonics-review}, low dissipation phonon transport \cite{phonon-review}, new type spintronics device application \cite{magnons-review} and so on.

\begin{figure*}[!tbp]
\centering\includegraphics[width=1\textwidth]{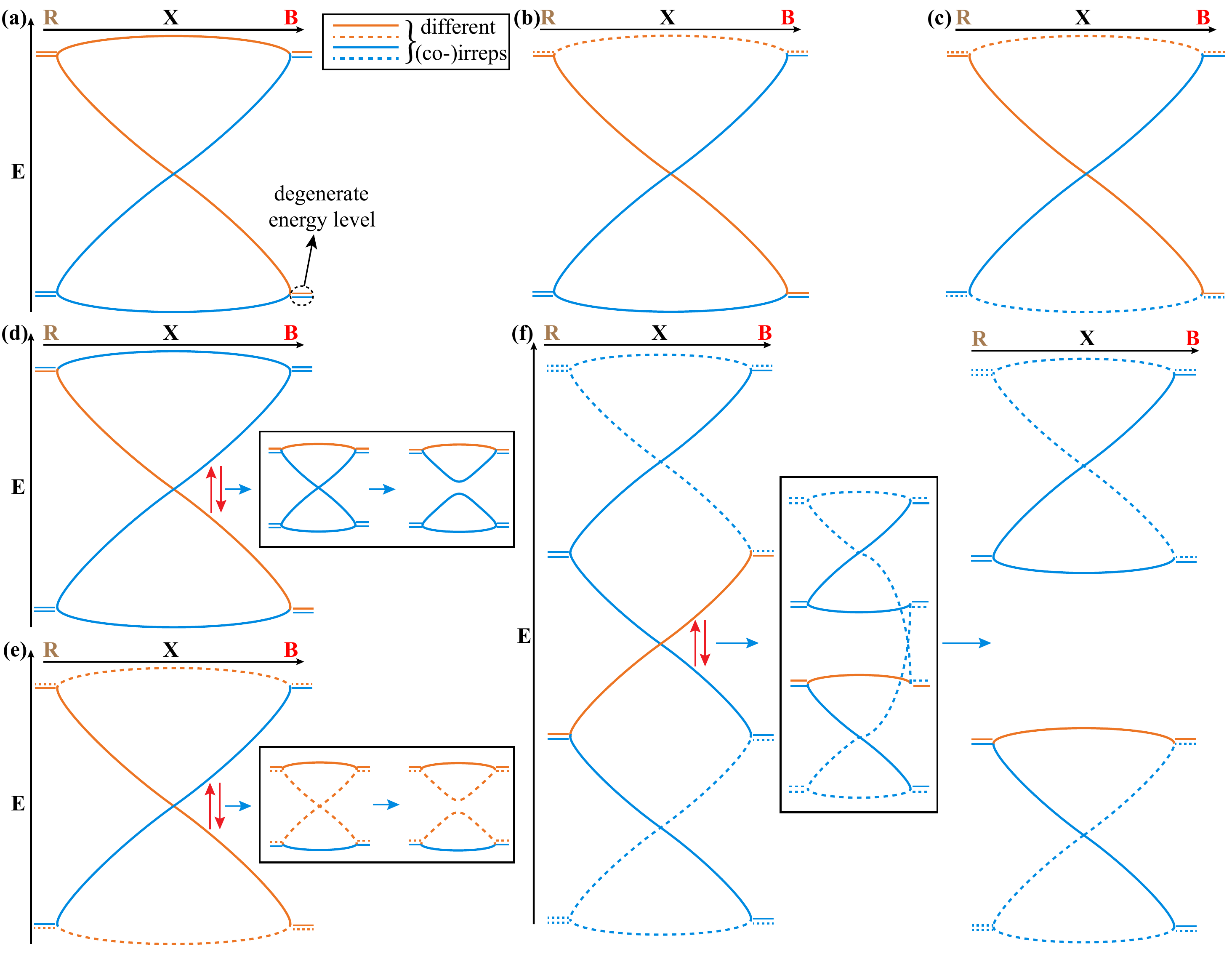}
\caption{Five types (A-E) of hourglass band structures formed along X which connects R and B, are schematically shown in: (a)-(e), respectively. We use different colors/styles of lines to represent different irreps or co-irreps (co-irrep is used when the little group contains antiunitary operation): Namely, the orange solid line, blue solid line, orange dashed line and blue dashed line represent four different irreps or co-irreps respectively.  Each energy level at R/B is represented by two horizontal bars, meaning that it is degenerate and would split to two sets of bands along X.  We use the same color and style for the horizontal bars as those for the lines representing the bands in X originated from the energy level, thus the band splitting pattern can be known easily. Note that (d) and (e) definitely correspond to nonessential hourglass band structure, since exchange of energy levels (e.g. at B here) as denoted by the up and down arrows would eliminate the hourglass shape (this is because the two bands with the same (co-)irrep would thus cross each other, and the resulting BC is not stable and gapped by nonvanishing hybridization not prohibited by symmetry, as shown in the insets). (a-c) might correspond to essential hourglass band structure, as described in detail in the main text. For them,  exchanging energy levels would not eliminate the hourglass shape (one can always find that bands with different colors cross each other). We also display connected hourglasses with the connecting point being an accidental BC, which could be gapped  by exchanging energy levels at B as shown in (f). The inverse is to create an accidental BC by connecting two hourglasses.}%, but note  another way  to make the hourglass band crossing nonessential is that there can exist another isolated band in X.}
\label{abcde}
\end{figure*}

Very recently, comprehensive classification of BCs for all the 230 space groups (SGs) \cite{Yao-1} or even 1651 magnetic space groups (MSGs) \cite{Tang-B-1,Tang-B-2, Yao-2, Yao-3} have been obtained. Especially, by Ref. \cite{Tang-B-2}, one can know both the associated low-energy $k\cdot p$ model and band nodal structures given a BC at some special high-symmetry $k$ point in the Brillouin zone (BZ), though whether the BC exists in a concrete material usually needs practical first-principles calculations \cite{N-1,N-2,N-3,N-4,Tang-NP, Tang-SA}. Symmetry-based classifications of band topology for all SGs or MSGs \cite{Po-NC, TQC-N, Haruki-SA, MTQC-NC, Slager-NP, Slager-X, Ono-SA, Ono-PRX} indicated that diverse topological phases can be formed in the spinful setting as well as the spinless setting \cite{LuLing-L-Haruki,LuLing-Res,Manes-B,CoSi-L,SE-chainphonon,SE-phonon-Wang,Rui-Hourglass,Rui-1,Rui-2,Rui-3,NS-1,NS-2,NS-3,phonon-NC,XuHu-L-1}.  Note that we usually require the BCs to be close to the Fermi level in electronic systems for observable phenomena, while there is no such requirement for bosonic systems. And some BCs  exist definitely (or essentially), for example, when the BC is located at a high-symmetry point, the irreducible representation (irrep) carried by the BC is the only possible irrep of the little group of the high-symmetry point. Another type of essential BC is due to nonsymmorphic symmetry enforcing special band connectivity such as hourglass one \cite{Hourglass-N, Hourglass-NC, ABO-L, ABO-B} (see Fig. \ref{abcde} for all types of hourglass band structures).  The essential BC is beneficial to searches for materials realizations of a target BC or complex nodal structures: For example, in the design of two nested nodal loops, one of which can already exist due to the essential  BC (the essential BC lies in the nodal loop), we only need to  make the other one. This is obvious much more easier than the method of making two nodal loops by tuning structure parameters \cite{Wu-B-2}. Such strategy of designing nodal structures might be easily implemented in artificial structures \cite{LuLing-NatPho,Chan-N, Chan-NC, LiFeng-NP, Deng-NC, Lu-NP, ZhangBaiLe-N, Qiu-PRL, Zhang-PRL, Jiang-N, Hughes-N, ChengJianChun-L, Sound-N, ChenYanFeng-NP,C-4-exp-1,C-4-exp-2,Dual-N,Lei-L,Lei-Hinge}.

In this work, we focus on all possible hourglass bosonic excitations in all the 1651 MSGs and 528 magnetic layer groups (MLGs), which can be obtained using compatibility relations (CRs) to enforce hourglass band connectivity. As applications, we perform first-principles calculations for hourglass phonons with time-reversal symmetry  in realistic three-dimensional (3D) materials \cite{ICSD} and theoretically-proposed two-dimensional (2D) materials \cite{2Ddata}, to which results of type II MSGs and MLGs are applied, respectively.  It is worth pointing out that our tabulation of MSGs or MLGs realizing hourglass bosonic excitations could be applied to designing artificial structures and tuning structure parameters as needed conveniently. Besides, results for type-I, III and IV MSGs/MLGs are applicable to systems without time-reversal symmetry, which can be of technological importance, for example, in realizing phonons with finite angular momentum\cite{Lifa-L-1,Lifa-L-2,ZhangX-S}. Besides, recently, the spin-space groups \cite{SSG-1,SSG-2,SSG-3} have received much attention which might be the symmetry of magnon Hamiltonian \cite{SSG-2}. Ref. \cite{SSG-2} shows that some spin-space groups could be isomorphic with some MSGs which are of type I, III or IV, thus the results for these MSGs could be applied to the corresponding cases to identify new-type topological properties of magnons \cite{NC-Feiye, KangkangLi-PRL, CuTeO-NP, CuTeO-NC, CuTeO-PRB}. %In phonons, classical mechanics and other bosonic systems, the physical mechanism of breaking time inversion has been proposed and verified in experiments.\cite{Wang_2015, PhysRevLett.115.104302, PhysRevX.5.031011, fleury2016floquet,PhysRevLett.114.114301,PhysRevLett.126.203601}.

%\vspace{1.7cm}
%In Sec. \ref{essential}, we overview the strategy of exhaustively listing all hourglass bosonic excitations and describe the results given in XX of the Supplementary Material, where the essential results are highlighted in red (one can also see Table \ref{table-1} in the main text listing only the names of MSGs/MLGs with essentially hourglass bosonic excitations). Besides, we find that several sets of hourglass bands can be connected by altering energy levels at $R$ or $B$ (defined in Sec. \label{essential}) to form an additional accidental band node. Then in Sec. \ref{material-3d} we describe the high-throughput calculations for hourglass phonons in 3D nonmagnetic materials and we choose AuX (X=I, Br)\cite{jagodzinski1959kristallstruktur,janssen1978crystal} both crystallized in SG 138 as illustrative examples and subsequent analyses of topological properties of the hourglass band crossings as well as surface state calculations. In Sec. \ref{material-2d}, we demonstrate that the ever-proposed 2D nonmagnetic material AlI\cite{zhou20192dmatpedia} could also host hourglass phonons.
%Finally, Sec. \ref{discussion} is devoted to Discussions and Perspectives.
%\section{The exhaustive list of MSGs and MLGs hosting essential hourglass band crossings}\label{essential}
%\vspace{1.7cm}
\section{Strategy}
We begin with briefly describing the strategy of exhaustively classifying and listing all hourglass BCs based on CRs. In this work, single-valued representations for the little groups of the MSGs are used, applicable for bosonic bands or electronic bands with spin-orbit coupling negligible. For electronic bands with spin-orbital coupling, double-valued representations should be adopted \cite{hu2022magnetic}. Interestingly, we find that, different from the hourglass BCs in the spin-orbital coupled electronic band structures, whose degeneracy can only take 2 and 4 \cite{hu2022magnetic}, the bosonic hourglass band crossings can be 2, 3 and 4-fold degenerate. In most cases, the degeneracy of the hourglass BC is 2. We list all the results of MSGs hosting 3-fold and 4-fold bosonic hourglass BCs in Table \ref{table-2}.  We should also point out that the so-called hourglass BC throughout this paper is the one located at the neck of the hourglass band structure.
\begin{table*} [!htp]
 \caption{List of 255 MSGs and 42 MLGs with  essential hourglass BCs.  The MSG is given in the form of X.Y in the  Belov-Neronova-Smirnova (BNS) notation \cite{bns} following Ref. \cite{Bradley} and the type of the MSG (I, II, III or IV) is given in the parentheses. Note that in some MSGs, we also give a, b or c in the parentheses, which means that the MSG allows layer structure and  the resulting MLG describes the layer perpendicular to the direction $\mathbf{a}$, $\mathbf{b}$ or $\mathbf{c}$, respectively. For each MSG/MLG in this table, there exist at least one hourglass band structure in some $k$ path (represented by X in Fig. \ref{abcde}).} \label{table-1}
\begin{tabular}{c|c|c|c|c|c|c|c|c|c}
\hline\hline
4.10(IV;b) &7.27(IV;c) & 7.29(IV;a)& 7.30(IV) & 13.70(IV;c)&13.74(IV)&14.80(IV)& 14.81(IV;a)& 14.82(IV;a)& 14.84(IV)\\
18.20(IV;c)&18.22(IV) &19.28(IV)&19.29(IV)&20.36(IV) &26.71(IV;b) & 26.72(IV;a)&26.75(IV;b)&26.76(IV)&26.77(IV) \\
27.83(IV;b)&27.84(IV)&27.85(IV)&27.86(IV) & 28.94(IV;a)&28.95(IV;a)&28.96(IV)&28.98(IV)&29.100(II;a)&29.101(III;a)\\
29.103(III;a)&29.104(IV)&29.105(IV;a)&29.106(IV;a)&29.107(IV;a)&29.108(IV)&29.109(IV)&29.110(IV)&30.116(IV)&30.118(IV) \\
30.120(IV) &30.121(IV) &30.122(IV) &31.128(IV)&31.129(IV)&31.133(IV)&32.139(IV) &32.140(IV;c)&32.142(IV) &32.143(IV) \\
33.145(II)&33.146(III) &33.148(III)&33.149(IV) &33.150(IV) &33.151(IV)& 33.152(IV) &33.153(IV) &33.154(IV) &33.155(IV)\\
34.160(IV) &34.161(IV) &34.162(IV) &36.178(IV)& 39.201(IV) &40.208(IV) &41.217(IV) &46.247(IV) &48.262(IV) &48.263(IV)\\
49.272(IV;b) &49.274(IV;b)& 49.275(IV) &49.276(IV) &50.284(IV) &50.285(IV) &50.286(IV) &50.288(IV) &51.299(IV;a) &51.301(IV;a)\\
51.303(IV) &51.304(IV) &52.305(I) &52.306(II) &52.308(III) &52.310(III) &52.311(III) &52.312(III) & 52.313(III) &52.314(IV)\\
52.315(IV) &52.316(IV) &52.317(IV) &52.318(IV) &52.319(IV) &52.320(IV) & 53.330(IV)&53.334(IV) &53.335(IV) &53.336(IV)\\
54.337(I;b) &54.338(II;b) &54.340(III;b) &54.342(III;b) & 54.343(III) &54.344(III;b) &54.346(IV;b) &54.347(IV) &54.348(IV;b) &54.349(IV)\\
54.350(IV;b) &54.351(IV) & 54.352(IV) &55.360(IV;c) &55.361(IV) &55.362(IV) &55.364(IV) &56.365(I) &56.366(II) &56.367(III)\\
56.369(III) &56.370(III) &56.372(IV) &56.373(IV) &56.374(IV) &56.375(IV) &56.376(IV) &57.377(I;c)&  57.378(II;c) &57.382(III;c)\\
57.383(III;c) &57.384(III) &57.385(III;c) &57.386(IV;c) &57.387(IV;c) &57.388(IV) & 57.389(IV) &57.390(IV) &57.391(IV;c) &57.392(IV)\\
58.400(IV) &58.402(IV) &59.412(IV) &59.414(IV)& 60.417(I) &60.418(II) &60.419(III) &60.422(III) &60.423(III) &60.424(III) \\
60.425(III) &60.426(IV) & 60.427(IV) &60.428(IV) &60.429(IV) &60.430(IV) &60.431(IV) &60.432(IV) &61.433(I) &61.434(II)\\
61.436(III) &61.437(III) &61.438(IV) &61.439(IV) &61.440(IV) &62.441(I) &62.442(II) &62.446(III) & 62.447(III) &62.448(III)\\
62.449(III) &62.450(IV) &62.451(IV) &62.452(IV) &62.453(IV) &62.454(IV) & 62.455(IV) &62.456(IV) &64.478(IV) &64.479(IV)\\
66.499(IV) &72.547(IV) &73.551(III) &76.11(IV) & 78.23(IV) &84.57(IV) &92.116(IV) &92.117(IV) &96.148(IV) &96.149(IV)\\
100.176(IV) &100.178(IV) & 101.185(IV) &101.186(IV) &102.192(IV) &103.201(IV) &103.202(IV) &104.208(IV) &106.221(III) &106.223(III)\\
106.224(IV) &106.226(IV) &110.249(III) &116.297(IV) &116.298(IV) &117.304(IV) &117.306(IV) &118.312(IV) & 124.361(IV) &124.362(IV)\\
125.372(IV) &125.374(IV) &126.384(IV) &126.385(IV) &127.396(IV) &127.398(IV) &130.423(I) &130.424(II) &130.426(III) &130.427(III) \\
130.429(III) &130.432(IV) &130.433(IV) &130.434(IV)& 131.445(IV) &132.457(IV) &132.458(IV) &133.464(III) &133.465(III) &133.468(IV) \\
133.470(IV) &134.480(IV) &134.481(IV) &135.483(I) &135.484(II) &135.486(III) &135.487(III) &135.488(III) &135.489(III) &135.492(IV) \\
135.493(IV) &135.494(IV) &138.519(I) &138.520(II) &138.522(III) &138.523(III) &138.525(III) &138.528(IV) & 138.529(IV) &138.530(IV)\\
142.567(III) &205.33(I) &205.34(II) &205.35(III) &205.36(IV) & & & & & \\
\hline
\end{tabular}
\end{table*}

\begin{table*}[!t]
\caption{List of MSGs which can host 3-fold and 4-fold degenerate bosonic hourglass BCs. The MSGs hosting essential hourglass band crossing are printed in bold. Note that for MLGs, only 2-fold degenerate hourglass bosonic BCs are allowed.}\label{table-2}
   \begin{tabular}{m{2.5cm}<{\centering}|m{15cm}<{\centering}}
    \hline
     \hline
    degeneracy &  MSGs \\\hline
    \makecell{3-fold degenerate \\ hourglass BC} & \makecell{215.73, 216.77, 218.82, 219.86, 220.90, 221.97, 222.99, 222.101, 222.102, 223.105, 223.108, 223.109, 224.115, \\225.121, 226.123, 226.126, 227.133, 228.135, 228.138, 230.146, 230.149} \\
\hline
    \makecell{4-fold degenerate \\ hourglass BC} & \textbf{52.315}, \textbf{54.352}, \textbf{57.388}, \textbf{60.428}, \textbf{62.452}, 125.374, 129.420, 132.458, 136.504, 137.516  \\
    \hline
  \end{tabular}
\end{table*}

As schematically shown in Fig. \ref{abcde}, we consider a trio denoted by R-X-B, which means that R and B are connected by X and definitely own higher symmetry than X. Then from R (B) to X, bands should split and the splitting pattern can be known based on CRs. As shown in Fig. \ref{abcde}, we use two horizontal bars to denote each degenerate energy level at R(B). The splitting pattern from the energy level at R(B) to X is also encoded in the colors/styles of the two horizontal bars representing the energy level. In Sec. IV of the  Supplementary Material \cite{supple-material}, we list all such trios allowing hourglass BCs. In Fig. \ref{abcde}, we use different colors/styles of lines, which represent energy bands in X, to denote different irreps or co-irreps of little group of X, carried by the bands in X. To form an hourglass shape of bands, two energy levels at R and B need to be considered, all of which split into two branches of bands along X. We can formally describe such splitting pattern by: X1$\oplus$X2,X3$\oplus$X4;X5$\oplus$X6,X7$\oplus$X8, namely, the higher (lower) energy level at R splits into two bands whose (co-)irreps are X1 and X2 (X3 and X4) while  the higher (lower) energy level at B splits into two bands whose (co-)irreps are X5 and X6 (X7 and X8). Due to the continuity of Bloch wave functions, we have \{X1,X2,X3,X4\}=\{X5,X6,X7,X8\}. Besides, to form an hourglass BC, \{X1,X2\}$\ne$\{X5,X6\} and  \{X3,X4\}$\ne$\{X7,X8\}.

Next, it should be required that X should allow at least two different (co-)irreps, thus X can be a high-symmetry line or high-symmetry plane. When X is a high-symmetry line, R and B should both be high-symmetry points and the hourglass BC can be a nodal point or lie in a nodal line within a high-symmetry plane \cite{Wu-B-2,Tang-B-2}.  When X is a high-symmetry plane, R and B can be high-symmetry point or line and the hourglass BC  definitely lies in a nodal line within X, in other word, each point in the nodal line is the BC point due to an hourglass band connectivity.  For the splitting pattern: X1$\oplus$X2,X3$\oplus$X4;X5$\oplus$X6,X7$\oplus$X8 which enforces an hourglass structure, we have X1=X5, X4=X8, \{X2,X3\}=\{X6,X7\} (but X2$\ne$X6). Hence, X2 and X3 are interchanged from R to B and thus constitute the (co-)irreps of the hourglass BC whose topological character can then be identified \cite{Tang-B-2}.  Based on the above requirements, we thus obtain five types of hourglass BCs: type A-E as shown in Fig. \ref{abcde}. Based on all (co-)irreps listed in Ref. \cite{Tang-B-1}, we calculate all CRs of all possible trios R-X-B and then list all possible hourglass BCs. They are all provided in Sec. IV of the Supplementary Material: For each MSG or MLG, the coordinates of R, X and B are given in the convention adopted in Ref. \cite{Bradley} alongside which the relevant band splitting patterns are also provided. The dimensions of relevant (co-)irreps of X are shown simultaneously, thus one can quickly know the degeneracy of the hourglass BC. The exhaustive list indicates that type-B hourglass bosonic excitations are very rare while type-A ones are the commonest cases.

Then we discuss more constraints for hourglass BCs to be essential, as printed in red in Sec. IV of the Supplementary Material.  As shown in Fig. \ref{abcde}, type D and E hourglass BCs cannot be essential, since exchange of energy levels could gap the hourglass BC. In Ref. \cite{Wu-B-1}, we focused only on the 230 SGs and obtained an exhaustive list of hourglass BCs and we also imposed a very strict condition for the hourglass BC to be essential there: it is required that the hourglass BC is of type A, B or C and the splitting pattern in the hourglass BC is the only possible splitting pattern.   However, here we find that for some MSGs, the hourglass BC can still  exist essentially in some $k$ path though there can be different splitting patterns in this path. All MSGs and MLGs with essential hourglass bosonic BCs are listed in Table \ref{table-1} and as shown by statistics in Table \ref{table-3}, many of MSGs/MLGs allowing hourglass BC can host essential hourglass BC. Here we say the splitting patterns are different once the symmetry contents (namely, the (co-)irreps in the hourglass band structure) are different.  Such interesting case is illustrated in Fig. \ref{abcde} (f) where we also show that an exchange of energy levels at R would give rise to another accidental BC connecting two hourglasses. A thorough check of our results reveals that only MSGs 134.481, 138.520 and 138.522 can host coexisting essential hourglass BCs of different types in some $k$ paths X, further found to include types-A, B and C. For other $k$ paths in these MSGs and for the rest MSGs and the MLGs in Table \ref{table-1}, the essential BCs in a $k$ path belong to only one type of hourglass band structure, but the splitting patterns could be different, listed below: %MSGs 134.481, 138.520, 138.522 could host type-A, B and C essential hourglass BCs (of this property);
MSGs  135.483, 135.484, 135.487, 135.493 could host type-C essential hourglass BCs (of this property); MSGs 26.72, 26.76, 27.86, 31.128, 31.133, 33.149, 34.161, 48.262, 55.360, 55.362, 56.372, 56.376, 58.400, 58.402, 62.450, 62.452, 62.453, 62.455, 106.221, 106.223, 131.445, 133.464, 133.465 and 135.488 could host type-A essential hourglass BCs; For MLGs based on MSGs 26.72 and 55.360 with the translation symmetry along $\mathbf{a}$ and $\mathbf{c}$ broken, respectively, type-A hourglass BCs could essentially exist in some $k$ path but their splitting patterns can be different. %Furthermore, we find that two type-B hourglasses can be connected by changing energy levels, though our comprehensive results show that the essential type-B hourglass BC can coexist with other type-A and type-C hourglass BCs.
Among these essential hourglass BCs with different splitting patterns, we find that MSGs 131.45, 134.481, 135.483, 135.484, 135.487, 135.493, 138.520 and 138.522  can host multiple essential hourglass band structures which can be tuned to be connected with each other, as shown in Fig. \ref{abcde}(f).

\begin{table}[!t]
\caption{Statistics of type-I/II/III/IV MSGs and MLGs with essential and nonessential (indicated in the parentheses after MSG or MLG in the first row) bosonic hourglass BCs: When an MSG or MLG of some type can host hourglass BC essentially, we count 1 for essential hourglass BC while for an MSG or MLG, it cannot host any essential hourglass BC,  we count 1 for nonessential hourglass BC. The counts for each case are given by numbers in bold.  We also selected 25 materials with good hourglass phonon band structures and the number in the parentheses denotes the SG number. }\label{table-3}
   \begin{tabular}{m{1.5cm}<{\centering}m{1.5cm}<{\centering}m{1.5cm}<{\centering}m{1.8cm}<{\centering}m{1.8cm}<{\centering}}
    \hline
    TYPE &  \makecell{MSG \\(essential)} & \makecell{MLG\\ (essential)} & \makecell{MSG\\(nonessential)} & \makecell{MLG \\(nonessential)}  \\\hline
    I & \textbf{11} & \textbf{2}  & \textbf{27} & \textbf{2} \\
    II & \textbf{13} & \textbf{3} & \textbf{64} & \textbf{8}  \\
    III & \textbf{49} & \textbf{8} & \textbf{130} & \textbf{18} \\
    IV & \textbf{182} & \textbf{29} & \textbf{146} & \textbf{8}   \\ \hline
    TOTAL & \textbf{255} & \textbf{42} & \textbf{367} & \textbf{36} \\
    \hline
    \hline
    \multicolumn{5}{c}{Selected Materials} \\
    \hline

TlF(57)  &  PtSi(62)  &  LaSi(62)  &  AsRh(62)  &  AuGa(62) \\
PdSi(62)  &  HgO(62)  &  TiSi(62))  &   HfSi(62)  &  SrZn(62) \\
 SrSi(62)  &  IrSi(62)  &  LuPt(62)  &  PtY(62)  &  ZrSi(62) \\
TiB(62) &  GeSe(62)  &   PW(62)  &   MoAs(62)   &  AgBa(62) \\
RhSi(62)  &   ZrTe(62)  &   GePd(62)  &  {\color{red}AuI(138)}  &  {\color{red}AuBr(138)} \\
    \hline
  \end{tabular}
\end{table}

\begin{figure*}[!t]
\centering\includegraphics[width=1\textwidth]{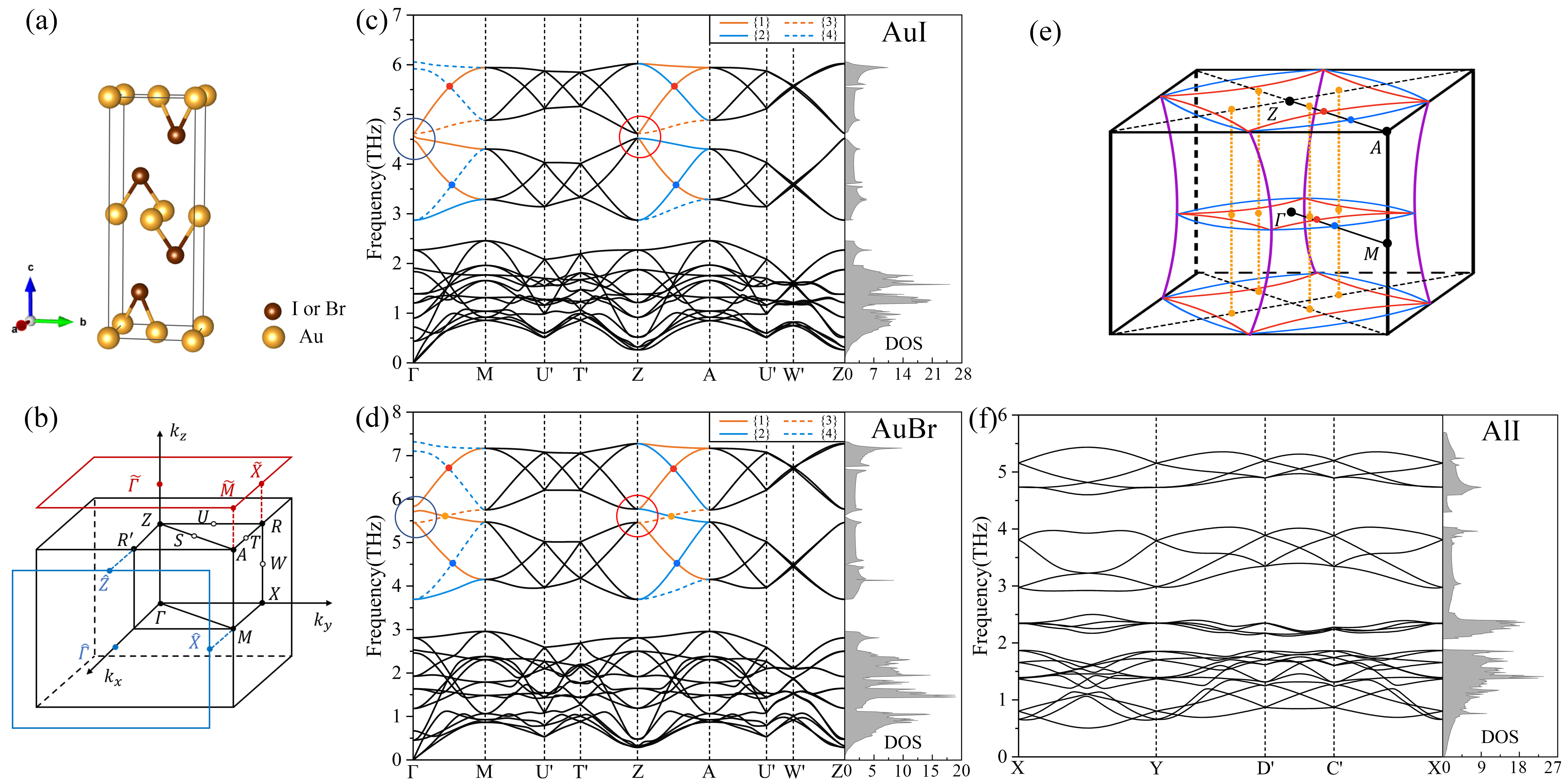}
\caption{ The crystal structures of AuI and AuBr in SG 138 are shown in (a) and their phonon band structures are shown in (c) and (d), respectively, where clear hourglass band structures can be observed around 3-7 THz. $U', T'$ and $W'$ are the middle point of the high-symmetry lines $U, T$ and $W$ shown in (b), respectively. Interestingly, (d) owns an additional BC which connects two nearby hourglasses due to the band switch of two branches in the red circle along $ZA$ high-symmetry line.  In (b), the BZ of SG 138 and (100) and (001) surface BZs are shown. (e) sketches the nodal lines emanating from the dots (red, orange and blue dots) in (d) of AuBr. (f) demonstrates the phonon band structures with clear hourglass bands in the proposed 2D material AuI. In (c,d,f), DOS means density of states per unit cell and per THz. In the insets of phonon spectra of (c) and (d), we indicate different co-irreps by different colors/styles of lines.}
\label{fig138}
\end{figure*}

\section{Materials studies}
 We then apply the results listed in Table \ref{table-1} to materials search. In the present work, we focus on type-II MSGs and MLGs to study the hourglass phonons in nonmagnetic 3D and 2D materials. Note that time-reversal breaking in phonon  might be realized in magnetic materials with strong spin-phonon coupling, and we thus leave the search or design of such materials with hourglass BCs in phonon spectrum to be a future work, for which the results of type-I/III/IV MSGs in Table \ref{table-1} still apply.  We also anticipate that the size of the hourglass band structure can be used as a smoking-gun of the strength of spin-phonon coupling. Hereafter we directly use  ``SG X'' to denote some type-II MSG named by X.Y in the BNS notation \cite{bns}. Then we are concerned with SGs 29, 33, 52, 54, 56, 57, 60-62, 130, 132, 135, 136, 138 and 205, among which SGs 29, 54, 57 also allow layer crystalline structures using Table \ref{table-1}. Note that SG 138 is very special since it can  host coexisting essential type-A, B and C hourglass BCs (within the high-symmetry line $ZA$ shown in Fig. \ref{fig138}(b)). It can also allow the connection of two different hourglass band structures. For 3D materials, we obtain 140 materials with the number of atoms $\le 10$ from the Inorganic Crystal Structure Database \cite{ICSD} and all these materials are found to be stable dynamically from their phonon spectra. We also find one 2D material proposed in the 2D materials database 2DMatPedia \cite{2Ddata}: AlI in SG 57 hosting hourglass BC whose phonon band structure is shown in Fig. \ref{fig138}(f). To our best knowledge, for 2D material, the hourglass band structure was only proposed in electronic band structure \cite{Hourglass-2D-L} and has never been proposed in the phonon spectrum before.   We select 25 good 3D materials candidates with clear hourglass band structures in Table \ref{table-3}.  We demonstrate all their phonon spectra along $k$-paths which essentially host hourglass BCs in Sec. III of Supplementary Material. In the following, we choose the 3D materials AuX (X=Br and I)\cite{AuI,AuBr} in SG 138 whose crystal structures are shown in Fig. \ref{fig138}(a) as the illustrative examples: Both materials host essential hourglass BCs in the phonon spectra, as required by the symmetry analysis. Interestingly, there is an additional BC in high-symmetry line ZA for AuBr, which is formed by connecting two hourglasses in the way demonstrated in Fig. \ref{abcde}(f). %The effect of such connection on nodal structures and surface states is also discussed.
\begin{figure*}[!tbhp]
\centering\includegraphics[width=1\textwidth]{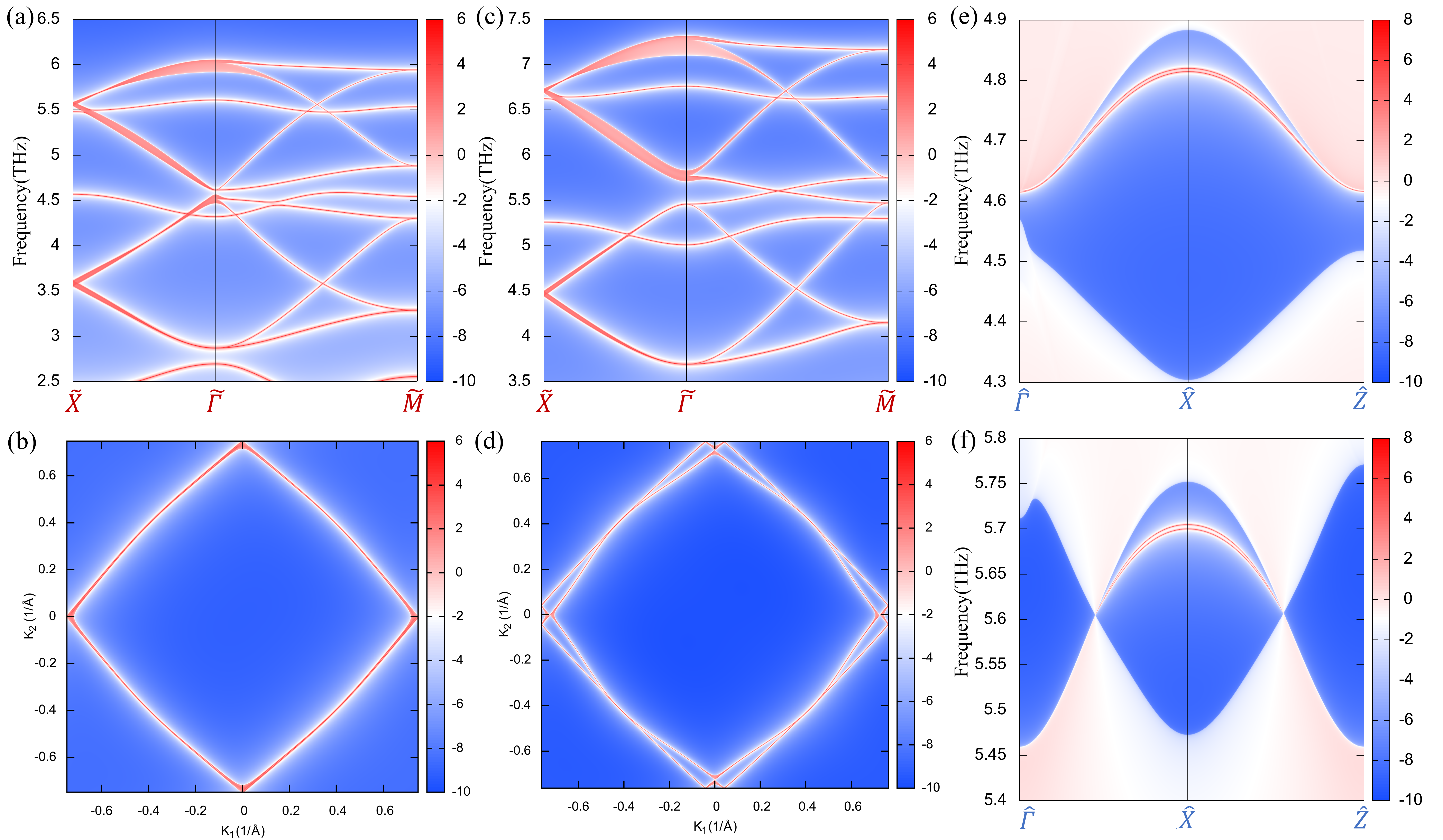}
\caption{The phonon band structures of the slabs with (001) surface terminated for AuI and AuBr are demonstrated in (a) and (c) respectively. The $k$ points of the corresponding surface BZ are depicted in Fig. \ref{fig138}(b). (b) and (d) shows the contour plots at the frequencies corresponding to the blue dots in Figs. \ref{fig138} (c) and (d)  of the (001) surface states for AuI and AuBr, respectively. (e) and (f) demonstrate the phonon band structures of the slabs with (100) surface terminated for AuI and AuBr, respectively. In Figs. 3 (a,c,e,f), the color denotes the local density of states at the surface along the chosen $k$ path within a frequency window. In Figs. 3(b,d), the color denotes the local density of states at the surface within the surface BZ for the frequencies at the blue dots in Figs. \ref{fig138}(c,d), respectively.}
\label{surf}
\end{figure*}
\section{Hourglass phonons in SG 138}
SG 138 with a tetrahedral lattice and $D_{4h}$ point group, owns multiple nonsymmorphic operations like 4-fold screw around $z$ axis, 2-fold screws around $x$ and $y$ axes and glides along $xy$, $yz$ and $xy$ planes, making this SG very special.  From Sec. IV.A of Supplementary Material, we know that high-symmetry line $S=(w,w,\frac{\pi}{c})$ ($ZA$ in Fig. \ref{fig138}(b)) and high-symmetry planes $P2=(u,v,\frac{\pi}{c})$ ($ZRAR'$ in Fig. \ref{fig138}(b)) and $P3=(0,u,v)$ ($\Gamma ZRX$ in Fig. \ref{fig138}(b)) can host hourglass BCs (the coordinates have been transformed into Cartesian ones). Note that though $S$ lies within $P2$, we should consider the hourglass band structures in $P2$ and $S$ separately, which could restrict the configuration of nodal line as shown later. $S, P2$ or $P3$, as the role of X described above, can be related by some high-symmetry points or lines, in the form of R-X-B listed below: $Z-S-A$, $Z-P2-A$, $Z-P2-T$, $A-P2-U$, $Z-P3-X$, $Z-P3-W$ and $U-P3-W$. $Z, A$ and $X$ are high-symmetry points whose coordinates are $(0,0,\frac{\pi}{c})$, $(\frac{\pi}{a},\frac{\pi}{a},\frac{\pi}{c})$ and $(0,\frac{\pi}{a},0)$, respectively. $T, U$ and $W$ are high-symmetry lines whose coordinates are $(w,\frac{\pi}{a},\frac{\pi}{c})$, $(0,w,\frac{\pi}{c})$ and $(0,\frac{\pi}{a},w)$, respectively.  We can further know that $S$ can host type-A, B, C essential hourglass BCs while $P2$ and $P3$ can only host type-A essential hourglass ones using Sec. IV.A of Supplementary Material. Since SG 138 owns spatial inversion symmetry, we should exploit co-irreps for all $k$ points in the form of $\{i\}$ or $\{i,j\}$ since they are all invariant under the combination of spatial inversion and time reversal operations \cite{Tang-B-1}. We find from Sec. IV.A of Supplementary Material that $S$ owns four different co-irreps denoted by $\{1\}$, $\{2\}$, $\{3\}$ and $\{4\}$, whose dimensions are all 1, namely, the bands along $S$ are all nondegenerate thus the hourglass BCs should be two-fold Weyl-like band nodes.  $P2$ and $P3$ both allow two different co-irreps $\{1\}$ and $\{2\}$, both of dimension 1. We show detailed analyses on the formation of hourglass BCs by CRs along $Z-S-A$ as follows while for the rest $k$ paths $P2$ and $P3$, similar process can apply. Consider the band splitting pattern only. From $Z$ to $S$, the splitting pattern can be $S\{1\}\oplus S\{3\}$, $S\{3\}\oplus S\{4\}$, $S\{2\}\oplus S\{4\}$, $S\{1\}\oplus S\{2\}$; From $A$ to $S$, the splitting pattern can be $S\{1\}\oplus S\{1\}$, $S\{2\}\oplus S\{2\}$, $S\{3\}\oplus S\{3\}$, $S\{4\}\oplus S\{4\}$, $S\{1\}\oplus S\{4\}$  and $S\{2\}\oplus S\{3\}$. It is then easy to find that bands in $S$ cannot be connected from $A$ to $Z$ pairwise and the least number of isolated branches in $S$ is 4, as the hourglass connectivity does. Through enumerating all possible combinations, we can know that the hourglass BCs can be composed of $S\{1\}\oplus S\{2\}$, $S\{1\}\oplus S\{3\}$, $S\{2\}\oplus S\{4\}$ or $S\{3\}\oplus S\{4\}$, while the other two possibilities of band nodes in $S$ own $S\{1\}\oplus S\{4\}$ and $S\{2\}\oplus S\{3\}$, which can appear accidentally. By first-principles calculations, we verify the hourglass phonons in $S$, $P2$ and $P3$ for both materials as shown in Figs. \ref{fig138}(c) and (d) for AuI and AuBr, respectively, demonstrating clear hourglass bands from  3  THz to 7 THz. From the density of states (DOS) in Fig. \ref{fig138}(c) and (d), those hourglass BCs are nearly ideal ones. For both materials, there are two essential hourglass BCs in $AZ$ composed of co-irreps $\{1\}\oplus\{2\}$ in this frequency window. Interestingly, for AuBr, an accidental BC due to the band switch between two frequencies with different co-irreps (indicated by a red circle in Figs. \ref{fig138}(c) and (d)) at $Z$ compared with those of AuI,  is formed, which connects two neighboring hourglasses. We also calculate the representations in $\Gamma-M$ of  the band branches related with these two hourglasses. The band switch at $\Gamma$ is also observed from the two frequencies in the  blue circles in Figs. \ref{fig138}(c) and (d), giving rise to an accidental BC for AuBr within $\Gamma-M$.

\section{Surface states}
We then compute the surface states in (001) and (100) surfaces for AuBr and AuI. Before that we firstly apply the results in Ref. \cite{Tang-B-2} to know the nodal structure originated from the BCs, denoted by orange, red and blue dots in Fig. \ref{fig138}(d) for AuBr: The red and blue dots should lie in nodal lines (in red and blue, respectively) in the horizontal planes ($ZRA$ and $\Gamma XM$ in Fig. \ref{fig138}(b)). The nodal lines in the two horizontal planes are found to be connected by four nodal lines in vertical planes (purple lines in Fig. \ref{fig138}(e)). Note that such purple lines connecting the red lines almost coincide with those for the blue lines. The red, blue and purple nodal lines also exist for AuI. For the orange dots for AuBr, they should lie in nodal lines in another two vertical planes (see the nearly straight orange lines in Fig. \ref{fig138}(e), which do not appear for AuI).

For the surface states originated from the nodal lines discussed above, we show the phonon band structures in Figs. \ref{surf}(a) and (c) for AuI and AuBr, respectively for the slab perpendicular to the (001) surface. Since $c$ is more than three times longer than $a$ and $b$, the phonons are almost not dispersive along $\mathbf{c}$ then we can observe hourglass shapes for the phonon band structures of the slab where the surface states are covered by the projection of bulk states. We then calculate the contour plots for frequency 3.58 THz and 4.53 THz of the (001) surface states for AuI and AuBr in Figs. \ref{surf}(b) and (d), respectively, found to be the projection of the nodal lines plotted in blue in the two horizontal planes in Fig \ref{fig138}(e). We also study the (100) surface states and the corresponding phonon bands for the corresponding slabs are shown in Figs. \ref{surf}(e) and (f) for AuI and AuBr, respectively, with the latter showing the appearance of  surface states emanating from the accidental BC (orange dot in $AZ$ of Fig. \ref{fig138}(d)).
\section{CONCLUSIONS and Perspectives}\label{discussion}
To conclude, we firstly list all essential and nonessential hourglass BCs in all the 1651 MSGs and 528 MLGs, using CRs calculated by the single-valued irreps or co-irreps listed in Ref. \cite{Tang-B-1}. The results for the 528 MLGs can be applied to 2D systems including 2D materials or surfaces of 3D materials. The property of the hourglass BC, such as the low-energy $k\cdot p$ model around it and the CR-required band splitting pattern,  can be further identified by looking up in Ref. \cite{Tang-B-2}. Based on the results for type-II MSGs, we calculate the phonon spectra of 140 3D materials which essentially host hourglass band BCs:  The hourglass BCs of some materials are at the intermediate frequency or high frequency, and is separated from other optical branches, which are thus easy to observe experimentally. The proposed 2D material AlI is also demonstrated to showcase hourglass phonons. 

We notice that Xia et al. experimentally demonstrated that there exist hourglass nodal lines in a photonic metacrystal at microwave frequencies \cite{photon-hourglass}. However, to the best of our knowledge, the hourglass band structure has not yet been observed in the phonon and magnon systems experimentally, but we believe our study could provide a basis for future fabrication of structures realizing the hourglass phonons and hourglass magnons easily.

The general bosonic excitation Hamiltonian should be restricted by  the MSG,  determined by the crystal and magnetic structure. However, for specific system, more symmetry could appear (that is to say, the MSG-allowed nonvanishing Hamiltonian parameter can be neglected). For example, for the phonon Hamiltonian of a magnetic material, though the material breaks time-reversal symmetry, the force constants are usually real-valued satisfying time-reversal symmetry, unless some mechanism (like spin-phonon coupling) enforces them to be complex-valued; In the magnon system, in some cases, the symmetry satisfies the spin space group, a supergroup of the MSG \cite{SSG-2}. In photonic systems, in addition to the symmetry of the MSG, it may own hidden symmetry of Maxwell equations \cite{xiong2020hidden}. Besides, in some artificial structures, duality can enhance the symmetry of Hamiltonian. It is possible to design metamaterials with emergent properties that escape standard group-theoretic analysis \cite{Dual-N}. Note that the  symmetry groups beyond the MSG symmetry, or the so-called non-crystallographic symmetry groups, have not yet been fully classified, which is a very interesting topic worthy of in-depth study in the future. However, our analysis is based on the most general Hamiltonian required by MSGs. For example, according to the MSG, it is found that a system subject to the MSG allows two-fold degeneracy, while including additional non-crystallographic symmetry, four-fold degeneracy might be allowed and doest not violate the requirement of two-fold degeneracy by MSG at all.  The formation of hourglass bosons based on non-crystallographic symmetry group is then left to be future work. Lastly, the hourglass band structure could be exploited to study the non-Abelian topology of real Hamiltonians \cite{WuQS-S,PRL-Z2,Slager-NC}.

\section{Acknowledgement}
We thank helpful discussions with Lin Wu, Zhipeng Cao and Yating Hu. F.T. was supported by National  Natural  Science Foundation of China (NSFC) under Grants No. 12104215. D.F., F.T. and X.W. were supported by the National Key R\&D Program of China (Grants No. 2018YFA0305704), NSFC Grants No. 12188101, No. 11834006, No. 51721001, and No. 11790311, and the excellent program at Nanjing University. X.W. also acknowledges the support from the Tencent Foundation through the XPLORER PRIZE.
\nocite{vasp-1,vasp-2,phonopy-1,vasp-pbesol,vasp-gga,vasp-paw,phonopy-non1,phonopy-non2,phonopy-non3,phonopy-sur1,phonopy-sur2}
\bibliography{references}
\end{document}